\documentclass{article}

\usepackage{arxiv}
\usepackage[utf8]{inputenc} 
\usepackage[T1]{fontenc}    
\usepackage{hyperref}       
\usepackage{url}            
\usepackage{booktabs}       
\usepackage{amsfonts}       
\usepackage{nicefrac}       
\usepackage{microtype}      
\usepackage{graphicx}
\usepackage{natbib}
\usepackage{doi}
\usepackage{color}
\usepackage{amsmath}
\usepackage{quantikz}
\usepackage{cite}
\usepackage{braket}
\usepackage{xcolor}
\usepackage{algorithm}
\usepackage{algpseudocode}
\usepackage{graphicx}
\usepackage{caption}
\usepackage{subcaption}
\usepackage{multirow}
\newcommand{\gb}{{\vec{\beta}, \vec{\gamma}}}

\usepackage[]{changes}
\definechangesauthor[name="Correction", color=red]{c}

\title{Graph Representation Learning for Parameter Transferability in Quantum Approximate Optimization Algorithm}

\author{ {Jose Falla} \\
	Department of Physics and Astronomy \\
	University of Delaware \\
	Newark, DE 19716 \\
	\texttt{jfalla@udel.edu} \\
	\And
	{Quinn Langfitt} \\
	Computational Science Division \\
	Argonne National Laboratory \\
	  Lemont, IL 60439 \\
        \texttt{qlangfitt@anl.gov} \\
        \And
        {Yuri Alexeev} \\
        Computational Science Division \\
        Argonne National Laboratory \\
        Lemont, IL 60439 \\
        \And
        {Ilya Safro} \\
        Department of Computer and Information Sciences \\
        Department of Physics and Astronomy \\
        University of Delaware \\
        Newark, DE 19716 \\
}

\hypersetup{
pdftitle={Graph Learning for Parameter Transferability in Quantum Approximate Optimization Algorithm},
pdfauthor={Jose Falla, Quinn Langfitt, Yuri Alexeev, Ilya Safro},
pdfkeywords={Quantum Computing, Quantum Approximate Optimization Algorithm, Quantum Optimization, Quantum Software, Parameter Transferability, Error Mitigation},
}

\begin{document}
\maketitle

\begin{abstract}
The quantum approximate optimization algorithm (QAOA) is one of the most promising candidates for achieving quantum advantage through quantum-enhanced combinatorial optimization. Optimal QAOA parameter concentration effects for special MaxCut problem instances have been observed, but a rigorous study of the subject is still lacking. Due to clustering of optimal QAOA parameters for MaxCut, successful parameter transferability between different MaxCut instances can be explained and predicted based on local properties of the graphs, including the type of subgraphs (lightcones) from which graphs are composed as well as the overall degree of nodes in the graph (parity). In this work, we apply five different graph embedding techniques to determine good donor candidates for parameter transferability, including parameter transferability between different classes of MaxCut instances. Using this technique, we effectively reduce the number of iterations required for parameter optimization, obtaining an approximate solution to the target problem with an order of magnitude speedup. This procedure also effectively removes the problem of encountering barren plateaus during the variational optimization of parameters. Additionally, our findings demonstrate that the transferred parameters maintain effectiveness when subjected to noise, supporting their use in real-world quantum applications. This work presents a framework for identifying classes of combinatorial optimization instances for which optimal donor candidates can be predicted such that QAOA can be substantially accelerated under both ideal and noisy conditions.
\end{abstract}


\section{Introduction}

Demonstrating quantum advantage in the mathematical optimization domain is poised to have a broad impact on science and humanity by allowing us to solve problems on a global scale, including finance~\citep{herman2022survey}, biology~\citep{outeiral2021prospects},  energy~\citep{joseph2023quantum} and scientific computing~\citep{Alexeev2021}. Variational quantum algorithms, a class of hybrid quantum-classical algorithms, are considered primary candidates for such tasks. These algorithms consist of parameterized quantum circuits, with parameters that are updated in classical computation. The quantum approximate optimization algorithm (QAOA)~\citep{farhi2014quantum} 
is a variational algorithm one of the applications of which is solving combinatorial optimization problems. In the domain of optimization on graphs, it has been demonstrated on such NP-hard problems as MaxCut~\citep{farhi2014quantum}, community detection~\citep{shaydulin2019network}, and partitioning~\citep{ushijima2021multilevel} by mapping these problems onto a classical spin-glass model (the Ising model) and minimizing the corresponding energy, a task that in itself is NP-hard.

Using machine learning (ML) methods has the potential to make significant breakthroughs in the field of quantum computing. Among many applications of ML in quantum computing, parameter prediction stands out as an important area, offering potentially significant improvement for executing quantum algorithms. For example, ML methods can be used to predict the ground state energy profile of a parametrized Hamiltonian\citep{cervera-lierta_meta-variational_2021} or generate a quantum state at an arbitrary point of a potential energy surface for molecules\citep{ceroni_generating_2023}.
The outcome of the quantum circuit execution is highly sensitive to the choice of parameters. However, the parameter space in quantum circuits is vast and often non-intuitive, making manual optimization or random guessing parameters a very inefficient process. This is where ML techniques, come into play. These methods are adept at navigating complex, high-dimensional spaces, identifying patterns and correlations that are not immediately apparent.

The application of ML in this context is not merely a matter of convenience but of necessity. Machine learning algorithms, through their adaptive learning capabilities, can systematically explore the parameter space, efficiently finding optimal parameters. This process potentially significantly improves the performance of quantum circuits by reducing the time and resources required for empirical trial-and-error methods or random guessing parameters. Moreover, the importance of ML-driven parameter optimization extends well beyond it. It is also an important factor in enhancing the resilience of quantum computations against errors.

It has been shown that by analyzing the distributions of subgraphs in two QAOA MaxCut instances one can predict how close the optimized QAOA parameters for one instance are to the optimal QAOA parameters for the other. From there, the concepts of donor (the graph which QAOA parameters will be reused) and acceptor (the graph that inherits QAOA parameters of donor to avoid costly optimization) graphs have been introduced. For example, by analyzing the overall parity (fraction of even-degree nodes) of both donor-acceptor pairs one can also predict good transferability between instances~\citep{galda_similarity-based_2023}. This prescription allows for successful transferability of QAOA parameters between two instances of the MaxCut problem, even in cases where the number of nodes of the acceptor instance is much greater than those of the donor instance. The measure of transferability of optimized parameters between QAOA instances on two MaxCut instances can be expressed with the approximation ratio, which is defined as the ratio of the energy of the corresponding QAOA circuit, evaluated with the optimized parameters $\gamma, \beta$, divided by the energy of the optimal MaxCut solution for the graph. While computing the optimal solution is hard in general case, for relatively small instances (graphs with up to 100 nodes are considered in this paper), it can be found by using classical algorithms, such as the Gurobi solver~\citep{gurobi}. 

\textbf{Our Contribution:} Building on these previous results, in this work we exploit structural graph features (subgraph composition) to apply graph representation learning models to a set of graphs generated with various node degree parity considerations, with the goal of predicting good donor candidates for target acceptor instances. We study the performance of five different whole graph embedding techniques by training the low dimensional representation of 30,000 40-node random graphs with pre-computed optimal parameters. We evaluate the performance of the graph embedding models for transferability for different classes of target graphs, including random, random regular graphs and Watts-Strogatz model graphs~\citep{watts1998collective}. The parameter optimization procedure for graphs in the training set is performed for a QAOA depth of $p = 3$, resulting in a set of 6 parameters to be transferred. The overarching idea of our parameter transferability pipeline is as follows: by using low dimensional representation of the test graph, we find its most similar pre-optimized graph and inherit its parameters.

We show that particular graph embedding models predict good donor candidates for the acceptor instances, even when the target acceptor is not a random graph. We also demonstrate that this finding significantly accelerates the QAOA solver by saving a lot of time in parameter optimization. For those graph embedding models that rely heavily on spectral features of the graphs (involving the adjacency and Laplacian matrices), good donor candidates are not always obtained, especially for the cases where the donor and acceptor instances are of in different graph model classes. 

We also demonstrate our transferability pipeline on the models with noise. More specifically, we show that the parameters associated with optimized donor graphs, as predicted by our graph embeddings, can still be effectively transferred to acceptor graphs on quantum processors that emulate the noise characteristics of IBM's Guadalupe and Auckland devices. Despite the noisy conditions, these parameters still manage to approach the ideal performance within acceptable error margins, affirming the potential of our approach to accelerate QAOA in the NISQ era.

This paper is structured as follows: In Section~\ref{sec:background} we present the relevant background material on QAOA, MaxCut, and graph embeddings. In Section~\ref{sec:graph_learning} we delve into the graph representation learning procedure for parameter transferability in QAOA. In Section~\ref{sec:results} we show the results of parameter transferability for 5 different graph embedding techniques, including results for noisy QAOA parameter optimization. In Section~\ref{sec:discussion} we provide a discussion about the different graph embedding models for parameter transferability and their use-case for different graph instances. Finally, In Section~\ref{sec:conclusions} we conclude with a summary of our results and an outlook on future advances with our approach.

\section{Background}\label{sec:background}

\subsection{Graph MaxCut problem}

The MaxCut problem, stemming from the Ising model that describes ferromagnetism within the context of statistical mechanics, constitutes a class of combinatorial optimization problem. The goal of the MaxCut problem is, given an unweighted, undirected simple graph $G = (V, E)$, to find a partition of the graph vertices $V$ into two disjoint sets such that the number of edges, $|E|$, between the two sets are maximized; or otherwise stated, to find a cut in the graph whose size is at least the size of any other cut. The MaxCut problem is known to be NP-Hard~\citep{WOEGINGER2005210}. 

\subsection{QAOA}

The QAOA is a hybrid quantum-classical algorithm that combines a parametrized quantum evolution with a classical outer-loop optimizer to approximately solve binary optimization problems \citep{farhi2014quantum}. The combinatorial optimization problem is defined on a space of binary strings of length $N$ and $m$ clauses. Each clause is a constrain satisfied by some assignment of the bit string. The objective function in this problem is defined as
 \begin{equation}
     C(z) = \sum_{\alpha=1}^{m}C_{\alpha}(z),
 \end{equation}
where $z = z_{1}z_{2}\cdots z_{N}$ is the bit string and $C_{\alpha}(z) = 1$ if $z$ satisfies the clause $\alpha$, and $0$ otherwise. QAOA maps the combinatorial optimization problem onto a $2^{N}$-dimensional Hilbert space with computational basis vectors $\ket{z}$ and encodes $C(z)$ as an operator $C$ diagonal in the computational basis. At each call to the quantum computer, a trial state is prepared by applying a sequence of alternating quantum operators
\begin{equation}
\label{eq:qaoa_state}
    \ket{\gb}_p := U_B(\beta_p)U_C(\gamma_p)\cdots U_B(\beta_1)U_C(\gamma_1)\ket{s}\,,
\end{equation}
where
\begin{equation}
    U_{C}(\gamma) = e^{-i\gamma C} = \prod_{\alpha=1}^{m}e^{-i\gamma C_{\alpha}}
\end{equation}
is the phase operator and
\begin{equation}
    U_{B}(\beta) = e^{-i\beta C} = \prod_{j=1}^{N}e^{-i\beta \sigma_{j}^{x}}
\end{equation}
is the mixing operator, with $B$ defined as the operator of all single-bit $\sigma^{x}$ operators, $B = \sum_{j=1}^{N}\sigma_{j}^{x}$. For the phase operator $\gamma \in (0, 2\pi)$ and for the mixing operator $\beta \in (0, \pi)$. The state $\ket{s}$ is some easy-to-prepare initial state, usually taken to be the uniform superposition product state
\begin{equation}
    \ket{s} = \frac{1}{\sqrt{2^{N}}}\sum_{z}\ket{z}.
\end{equation}
The parametrized quantum circuit (\ref{eq:qaoa_state}) is called the QAOA \textit{ansatz}, the number $p$ of alternating phase and mixing operators is the \textit{depth}, and the selected parameters $\gb$ define the \textit{schedule}, analogous to quantum annealing.

\begin{figure}[h!]
    \centering
    \includegraphics[width=0.8\linewidth]{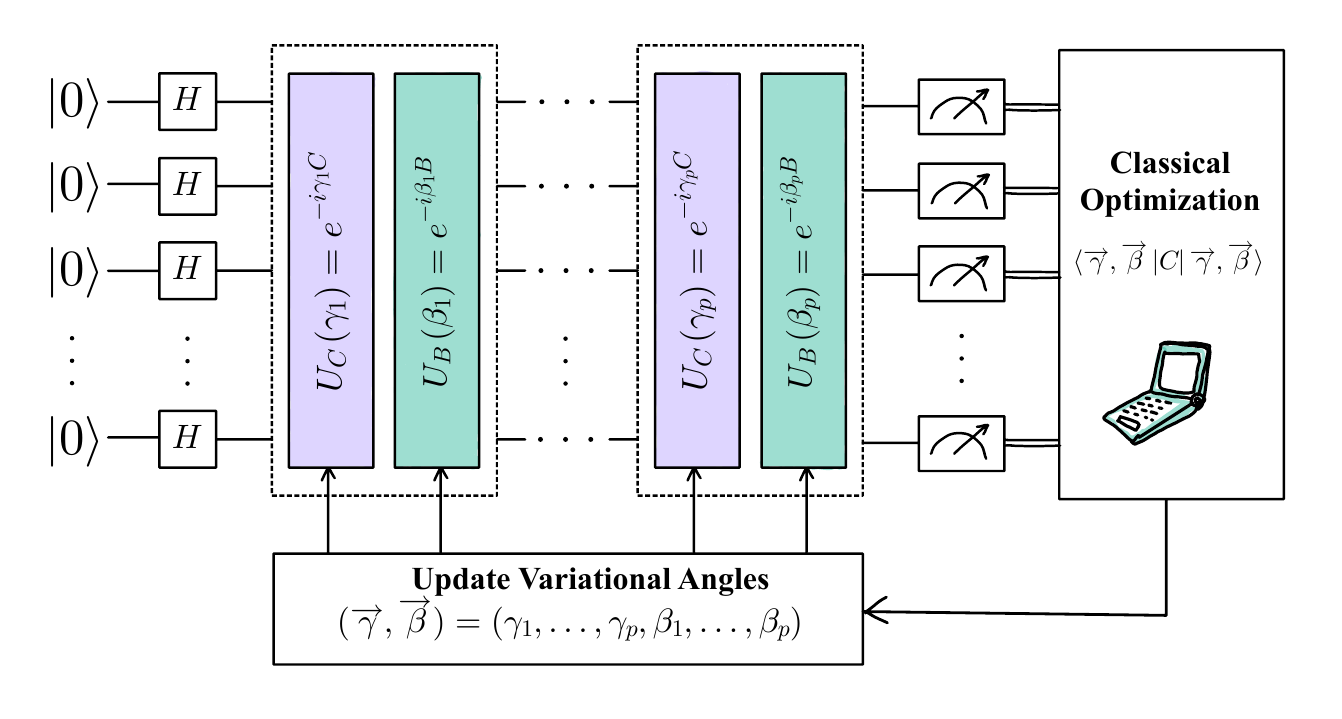}
    \caption{Schematic pipeline of a QAOA circuit. A parametrized ansatz is initialized, followed by series of applied unitaries that define the depth of the circuit. Finally, measurements are made in the computational basis, and the variational angles are classically optimized. This hybrid quantum-classical loop continues until convergence to an approximate solution is achieved.}
    \label{fig:qaoa_schem}
\end{figure}

Preparation of the state (\ref{eq:qaoa_state}) is followed by a measurement in the computational basis. The output of repeated state preparation and measurement is then used by a classical outer-loop optimizer to select the schedule $\gb$, based on the optimization of the expectation value of the objective function
\begin{equation}
    \langle C\rangle_p = \bra{\vec{\beta}, \vec{\gamma}}C\ket{\vec{\beta}, \vec{\gamma}}.
\end{equation}
The output of the overall procedure is the best bit string $z$ found for the given combinatorial optimization problem. Figure \ref{fig:qaoa_schem} shows a schematic of the QAOA procedure.

In the context of the MaxCut problem, the QAOA input is a graph with $|V| = N$ vertices and an edge set $\{\langle ij \rangle\}$ of size $m$, and the goal is to find a bit string $z$ that maximizes:
\begin{equation}
    C = \sum_{\langle ij \rangle}C_{\langle ij \rangle},
\end{equation}
where
\begin{equation}
\label{eq:cost_function}
    C_{\langle ij \rangle} = \frac{1}{2}(-\sigma_{i}^{z}\sigma_{j}^{z} + 1).
\end{equation}
It has been shown that for QAOA with depth $p = 1$ on a 3-regular graph MaxCut instance produces a solution with an approximation ratio of at least 0.6924 \citep{farhi2014quantum}, where the approximation ratio is defined as:
\begin{equation}
    r^{*} = \frac{\bra{\vec{\beta}, \vec{\gamma}}C\ket{\vec{\beta}, \vec{\gamma}}}{C^{*}} = \frac{\langle C\rangle_p}{C^{*}},
\end{equation}
where $C^{*}$ is the classical solution. The result by Farhi has been extended to include lower bounds to the approximation ratio in 3-regular graphs, where lower bounds of 0.7559 and 0.7924 have been found for QAOA depths of $p = 2$ and $p = 3$, respectively~\citep{wurtz_maxcut_2021}. In the limit $p \rightarrow \infty$, the classically optimal solution is achieved (i.e. $r^{*} \rightarrow 1$).
\subsection{Graph Embedding}
Graph embedding methods is a class of approaches used to transform nodes, edges, graphs and their features into vectors in low-dimensional vector space in a such way that aforementioned objects that exhibit common structural properties are close to each other in this vector space with respect to some distance function. In our case, we refer to graph embeddings that learn a mapping of the entire graph (i.e., a whole graph is a single data point) to a low-dimensional vector in some space space while preserving particular structural properties of the graph. Graph embedding techniques have shown remarkable capacity of converting high-dimensional sparse graphs into low-dimensional, dense and continuous vector spaces~\citep{cai_comprehensive_2018, goyal_graph_2018}. These non-linear and highly informative graph embeddings in the latent space can be used to address different downstream graph analytic tasks, such as node classification, link prediction, and community detection, to name a few.

While many graph embedding techniques focus on node embedding~\citep{sybrandt2020fobe,ding2020unsupervised, rozemberczki_fast_2020,yang_nodesketch_2019, grover_node2vec_2016} at various scales (microscopic, mesoscopic, and macroscopic node embedding), whole graph embedding techniques have also emerged to as representation learning methods for analyzing whole networks~\citep{wang_graph_2021, cai_simple_2022, galland_invariant_2019, narayanan_Graph2Vec_2017}. Whole graph embedding techniques can allow us to determine whether two graphs are structurally similar. A typical application of this approach involves classifying graphs based on their similarity.

Methods for graph classification\footnote{Intuitively, we treat our parameter transferability problem as a graph similarity problem.} can be grouped into a few categories. Among these categories, a classic family of methods is that involving graph kernels, with examples including the Weisfeiler-Lehman kernel~\citep{shervashidze_weisfeiler-lehman_2011}, random walk kernel~\citep{gartner_kernels_2003}, shortest path kernel~\citep{borgwardt_shortest-path_2005}, and deep graph kernel~\citep{yanardag_deep_2015}. Another family of methods is that involving graph embeddings for learning vector representations of graphs. This family of methods include Graph2Vec~\citep{narayanan_Graph2Vec_2017}, which uses first the Weisfeiler-Lehman kernel to extract rooted subgraph features that are then passed to a Doc2Vec~\citep{quoc_doc2vec_2014} model to get embeddings; also, GL2Vec~\citep{chen_gl2vec_2019}, which is an extension of Graph2Vec that includes line graphs to account for edge features. Other methods include SF~\citep{de_lara_simple_2018}, NetLSD~\citep{tsitsulin_netlsd_2018}, and FGSD~\citep{verma_NIPS_2017}, that use the information from Laplacian matrix and eigenvalues of a graph to generate embeddings. The Geo-Scatter~\citep{gao_geometric_2019} and FEATHER~\citep{rozemberczki_characteristic_2020} methods employ normalized adjacency matrices to capture the probability distribution of neighborhoods in graphs.

\section{Related Work}

The task of finding good QAOA parameters is challenging in general, for example, because of such reasons as encountering barren plateaus~\citep{anschuetz_beyond_2022, wang_noise-induced_2021}. Furthermore, the approximation ratio increases only marginally as the depth of the QAOA circuit is increased, and the gains are offset by the increasing complexity of optimizing variational parameters~\citep{shaydulin_2019_eval}. Acceleration of optimal parameter search for a given QAOA depth $p$ can be either incorporated into or be the main focus of many approaches aimed at demonstrating quantum advantage. Examples include  warm- and multi-start  optimization~\citep{egger2020warmstarting,shaydulin2019multistart}, problem decomposition~\citep{shaydulin2019hybrid},  instance structure analysis~\citep{shaydulin2021classical}, and parameter learning~\citep{khairy2020learning}. Optimal QAOA parameter transferability has shown great promise in circumventing the problem of finding good QAOA parameters~\citep{galda_similarity-based_2023,galda2021transferability}. Based on structural graph features, successful parameter transferability can be achieved between a donor instance that is much smaller, therefore easier to optimize parameters for, than the acceptor instance. As of the writing of this article, we are unaware of graph embedding techniques being used as models to determine graph donor candidates for optimal QAOA parameter transferability.

\section{Graph Learning Model for Parameter Transferability}\label{sec:graph_learning}

Graph  models for learning graph representation can greatly improve the computational cost of QAOA parameter optimization for particular MaxCut instances, especially for target instances that are large and have a complex connectivity and the depth of the circuit $p$ is increased (as the computational cost grows rapidly with the circuit depth), by employing model that is trained on graphs whose optimal QAOA parameter are known \textit{a priori}. By employing parameter transferability, the solution (optimal $\Vec{\gamma}$ and $\Vec{\beta}$ parameters) to a different (usually smaller) graph instance are transferred to the target instance and its QAOA energy is evaluated. The optimal donor parameters can be either applied directly to the acceptor state's construction, or used as a ``warm start'' for further optimization. The problem then becomes finding suitable donor candidates for particular target instances.

\begin{figure}[h!]
    \centering
    \includegraphics[width=0.8\linewidth]{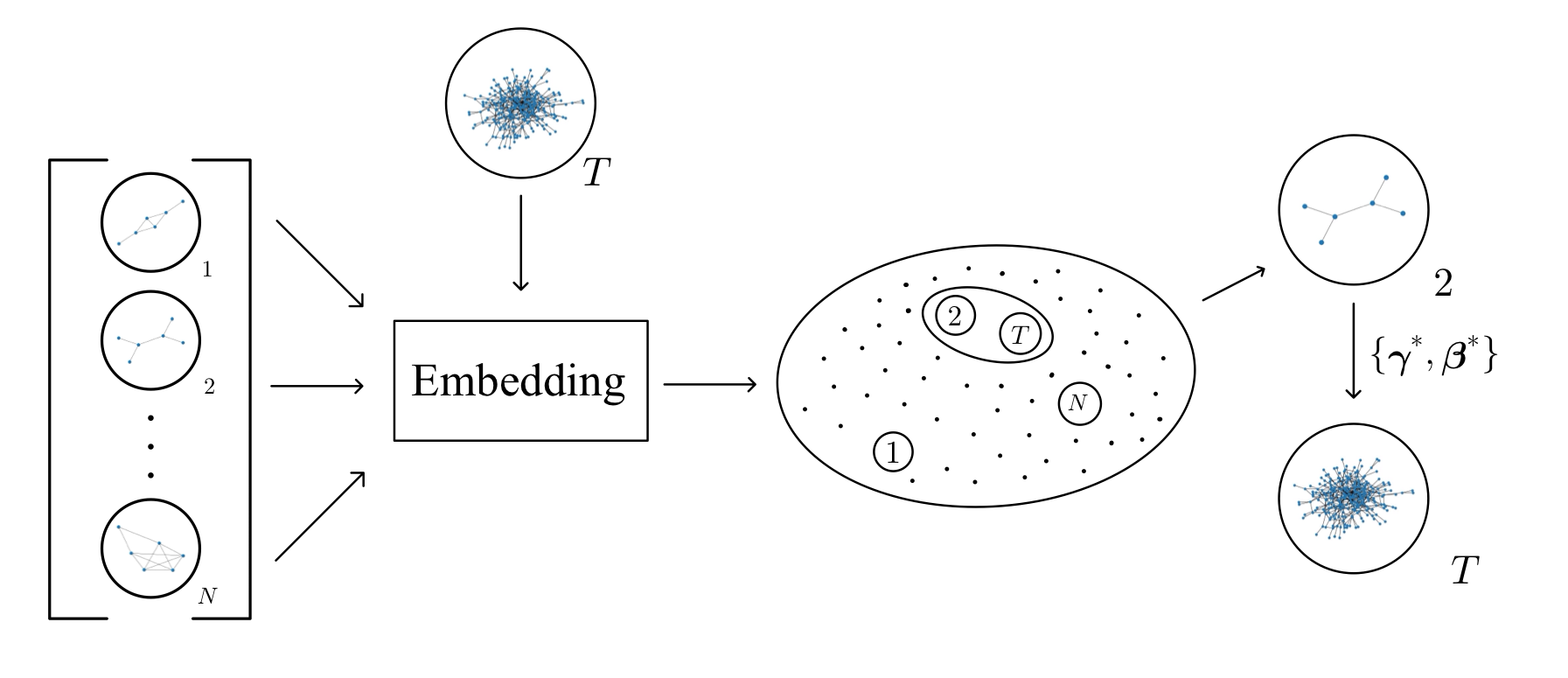}
    \caption{Schematic pipeline of embedding procedure. A list of graphs is prepared with parity considerations and the model is trained. A target acceptor graph (T) is projected into the embedded space. The best donor graph candidate is chosen by calculating all Euclidean distances between the target acceptor's embedded vector and the graphs in the training set and choosing the shortest distance. Finally, optimal QAOA parameters are transferred from the closest graph in the training set (optimal donor) to the target acceptor graph, and the target's QAOA energy is evaluated.}
    \label{fig:embedding_schem}
\end{figure}

Optimal QAOA parameter concentration effects have been reported for several types of graphs on which the MaxCut is formulated, mainly those involving random 3-regular graphs~\citep{brandao2018fixed, streif2020training, akshay2021parameter}. Brandao \textit{et al.}~\citep{brandao2018fixed} observed that the optimal QAOA parameters on a MaxCut problem for a 3-regular graph are also nearly optimal for all other 3-regular graphs. There are three possible subgraphs of 3-regular graphs; with one of these subgraph types being a tree. The authors note that in the limit of large $N$ (the number of nodes), the fraction of tree graphs asymptotically approaches 1, while the other two types of 3-regular subgraphs are of order $1 / N$, rendering almost all edges' neighborhoods locally to look like trees. Therefore, regardless of the parameter values, the objective function is the same for almost all 3-regular graphs, up to order $1/N$.

Based on the results of~\citep{galda_similarity-based_2023}, there are two structural graph features that predict good transferability between two MaxCut instances. The first feature is subgraph composition, where a similarity between instances is established by the number of shared (isomorphic) subgraphs. As each of the subgraphs contribute, in aggregate, to the MaxCut energy, so do the contributions from transferred subgraphs to a different MaxCut instance. Therefore, if two MaxCut instances possess a similarity in subgraph composition, it is highly likely that optimal QAOA parameters will transfer well.

The second feature, closely related to the first, that predicts good transferability is the parity of the graph. The parity of a graph $G(V,E)$ is defined as the fraction of nodes in $G$ with an even degree:
\begin{equation}
    \label{eq:parity}
    \pi_{G} = \frac{n_{even}}{|V|},
\end{equation}
where $n_{even}$ is the number of even-degree nodes, and $|V| = N$ is the total number of nodes. For the case of 3-regular graphs in Brandao \textit{et al.}~\citep{brandao2018fixed}, all graphs share the same parity and have good parameter transferability amongst them. This is extended to families of odd- and even-regular graphs, where good parameter transferability is achieved between families of odd- and even-regular graphs, and poor transferability between an odd-regular donor and an even-regular graph, and vice versa~\citep{galda_similarity-based_2023}. This work has shown that the use of parity can be extended to predict good transferability among different types of graphs, such as general random graphs.

Since a correlation has been established between structural graph features (subgraph composition and parity) and successful transferability of optimal QAOA parameters, we apply a set of graph embedding models that exploit these features to predict good donor candidates for target acceptor graphs. In particular, we focus on some previously mentioned models, namely: Graph2Vec \citep{narayanan_Graph2Vec_2017}, GL2Vec~\citep{chen_gl2vec_2019}, wavelet characteristic~\citep{wang_graph_2021}, SF~\citep{de_lara_simple_2018}, and FEATHER~\citep{rozemberczki_characteristic_2020}.

To this end, we construct a set of 30,000 non-isomorphic 40-node random graphs as a training set for all the graph embedding models. We test the performance of these models by computing approximation ratios (will be explained in Eq. (\ref{eq:avg_approx_ratio})) from the QAOA energy in the acceptor instance evaluated with the predicted donor's optimal parameters and the classical MaxCut energy of the acceptor instance.

\subsection{Learning Procedure}

\subsubsection{Graph2Vec}

We begin with the Graph2Vec~\citep{narayanan_Graph2Vec_2017} learning procedure, which enables graph representation as fixed-length feature vectors. The advantages of using this neural embedding framework is that Graph2Vec learns graph embedding in a completely unsupervised  manner (there is no need for class labels), that the embeddings are generic (task-agnostic), that it learns graph embeddings from a large corpus of graph data (unlike graph kernels, where features are handcrafted), and that it samples and considers rooted subgraphs, which ensures that the representation learning process yields similar embeddings for structurally similar graphs~\citep{yanardag_deep_2015, shervashidze_weisfeiler-lehman_2011}.

The learning procedures begins by constructing a training set of graphs $\mathbb{G}=\{G_{1}, G_{2}, ..., G_{n}\}$. First, the feature vector $\Phi$ is initialized by sampling $\Phi$ from $\mathbb{R}^{|\mathbb{G}|\times\delta}$, where $\delta$ is the embedding dimensionality. For each graph $G_{i}$, a rooted subgraph $sg_{j}^{(d)}$ of maximum node degree $d$ is extracted around every node $j$. The set of all rooted subgraphs produce a vocabulary $SG_{\textrm{vocab}} = \{sg_{1}^{(d)}, sg_{2}^{(d)}, ...\}$. The feature vector is updated as:
\begin{equation}
    \label{eq:cost}
    \Phi = \Phi - \alpha\frac{\partial J}{\partial\Phi},
\end{equation}
where $\alpha$ is the learning rate, and $J(\Phi) = -\log{Pr(sg_{j}^{(d)}|\Phi(G))}$ is the cost function: the log-likelihood probability of finding subgraph $sg_{j}^{(d)}$ in the context of the graph embedding $\Phi(G)$. Stochastic gradient descent is used to optimize the parameters in Equation \ref{eq:cost}. The learning procedure happens under $\mathfrak{e}$ epochs, and both the learning rate $\alpha$ and the rooted subgraph degree $d$ are empirically tuned. Given that the cost function depends on the log-likelihood probability of finding a subgraph within the context of other subgraphs, Graph2Vec ensures that the low-dimensional vector representation of a graph preserves the graph's structure, making it an ideal candidate for determining structural similarities between two graph instances. As mentioned previously, subgraph similarity is a key factor that determines good transferability of optimal parameters between two graph instances, thus, Graph2Vec can be readily applied to this end.

To demonstrate the use of Graph2Vec in the context of parameter transferability, we construct a training set of 30,000 40-node random graphs. For each of these graphs, 20 QAOA optimization procedures for MaxCut are performed, with a QAOA depth of $p = 3$. For a depth of $p = 3$, we obtain a set of $2p$ optimal parameters: 3 optimal $\gamma$ parameters and 3 optimal $\beta$ parameters. That is, for each of the graphs in the training set, there are 120 associated optimal parameters for all 20 optimizations. The graph set is constructed with the following parity considerations: there are 1,400 graphs for each of the possible 21 graph parities in 40-node random graphs. That is, we start with all 40 nodes having odd degree in the first 1,400 graphs, then make two out of the 40 nodes even degree in the next 1,400 graphs, and continue in this fashion until the last 1,400 graphs have all 40 nodes with even degree. Furthermore, the maximum node degree is limited to $d = 4$.

For the learning procedure, we use the unsupervised machine learning extension for NetworkX, Karate Club package \citep{karateclub}. The training is performed with the default embedding dimensions (128), number of workers (4), number of feature extraction recursions (2), and down sampling rate for frequent features (0.0001). Otherwise, the training is performed over $\mathfrak{e}=100$ epochs at a learning rate of $\alpha = 0.065$. The validity of the model is checked by performing the graph embedding and testing with the same training set. That is, the training set is used as a test set by projecting these graphs again into the embedded space and finding the Euclidean distance between the test set's and training set's feature vectors. For each of the graphs, if the Euclidean distance is zero, the model correctly predicts the most similar graphs in terms of their rooted subgraphs. Our model correctly predicts $\sim$98\% of the test set against the training set.

After the learning procedure is conducted, a test (acceptor) graph is passed through the model and projected on to the embedded space. The test graph's feature vector is compared against the training set graphs' feature vectors and the training set graph that produces the minimum Euclidean distance is selected as the optimal donor, with the Euclidean distance being measured as:
\begin{equation}
    d(\Vec{p}, \Vec{q}) = \sqrt{\sum_{i = 1}^{n}(q_{i} - p_{i})^{2}},
\end{equation}
where $p$ and $q$ are two point in Euclidean $n$-space. After the optimal donor is selected, all 20 sets of optimal parameters are transferred to the acceptor graph, and the average transferred approximation ratio is computed as:
\begin{equation}
    \label{eq:avg_approx_ratio}
    r_{\textrm{avg}} = \frac{1}{20}\sum_{i = 1}^{20}\frac{A(\Vec{\gamma}_{D_{i}},\Vec{\beta}_{D_{i}})}{A^{*}},
\end{equation}
where $A(\Vec{\gamma}_{D_{i}},\Vec{\beta}_{D_{i}})$ is the QAOA energy of the acceptor graph evaluated with the donor graph's transferred parameters, and $A^{*}$ is the classical optimal energy for acceptor graph $A$\footnote{In general, for large instances of the MaxCut problem, a classical solution becomes prohibitively expensive to compute and the true approximation ratio is rendered an unobtainable metric.}. The classical MaxCut energy is computed using the Gurobi solver\footnote{The Gurobi solver provides  classically optimal MaxCut solutions in a competitive speed with known optimization gap. For the purpose of this work, there is no particular reason to choose Gurobi over IPOPT or other similarly performing solvers.}. The QAOA energy is computed using QTensor~\citep{qtensor}, a large-scale quantum circuit simulator based on a tensor network approach, thus, it can provide an efficient approximation to certain classes of quantum states~\citep{kardashin_quantum_2021, biamonte_tensor_2017}.

\subsubsection{Learning Procedure for Other Models}

We explicitly present the learning procedure for the Graph2Vec algorithm, as it is the algorithm that performs better predictions overall (refer to Section \ref{sec:results}). For the learning procedure of all other algorithms, we refer the readers to: GL2Vec~\citep{chen_gl2vec_2019}, wavelet characteristic~\citep{wang_graph_2021}, SF~\citep{de_lara_simple_2018}, and FEATHER~\citep{rozemberczki_characteristic_2020}\footnote{All graph embedding algorithms employed in this work are unsupervised learning models.} As an alternative to the embedding, additional transerability strategy is based on solving the graph alignment problem \citep{elruna}.

\section{Computational Results}\label{sec:results}

\subsection{Experimental Setting}

For each of the graphs in the training set that contains $\sim 30,000$ random graphs, 20 optimization runs were performed using the tensor network simulator QTensor~\citep{qtensor} for a QAOA depth of $p = 3$. Therefore, for each graph in the training set, there are 60 associated optimal parameters.

For the training of the graph embedding models, we use KarateClub's implementations of the embeddings~\citep{karateclub}. The structural feature models Graph2Vec and GL2Vec are trained with the number of epochs set to $\mathfrak{e}=100$, and a learning rate of 0.065, with all other hyperparameters set at default. For the spectral feature learning embeddings (SF and Wavelet Characteristic), all models are trained with the default settings. Default settings are also used for the random-walk-based model FEATHER.

Finally, for the ELRUNA method, the symmetric substructure score ($S^{3}$) was calculated between the target acceptor graph and all graphs in the training set, with seed alignment.

After training the appropriate model, optimal parameter transferability was performed using test graphs that consist of random graphs, regular graphs, and Watts-Strogatz graphs. Therefore, optimal parameter transferability is not only performed between the same type of graphs, but also between different types of graphs.

\subsection{Results Without Noise}

\subsubsection{Graph2Vec}

We begin by looking at parameter transferability with the structural graph feature learning model Graph2Vec.

\begin{figure}[h!]
    \centering
    \includegraphics[scale=0.25]{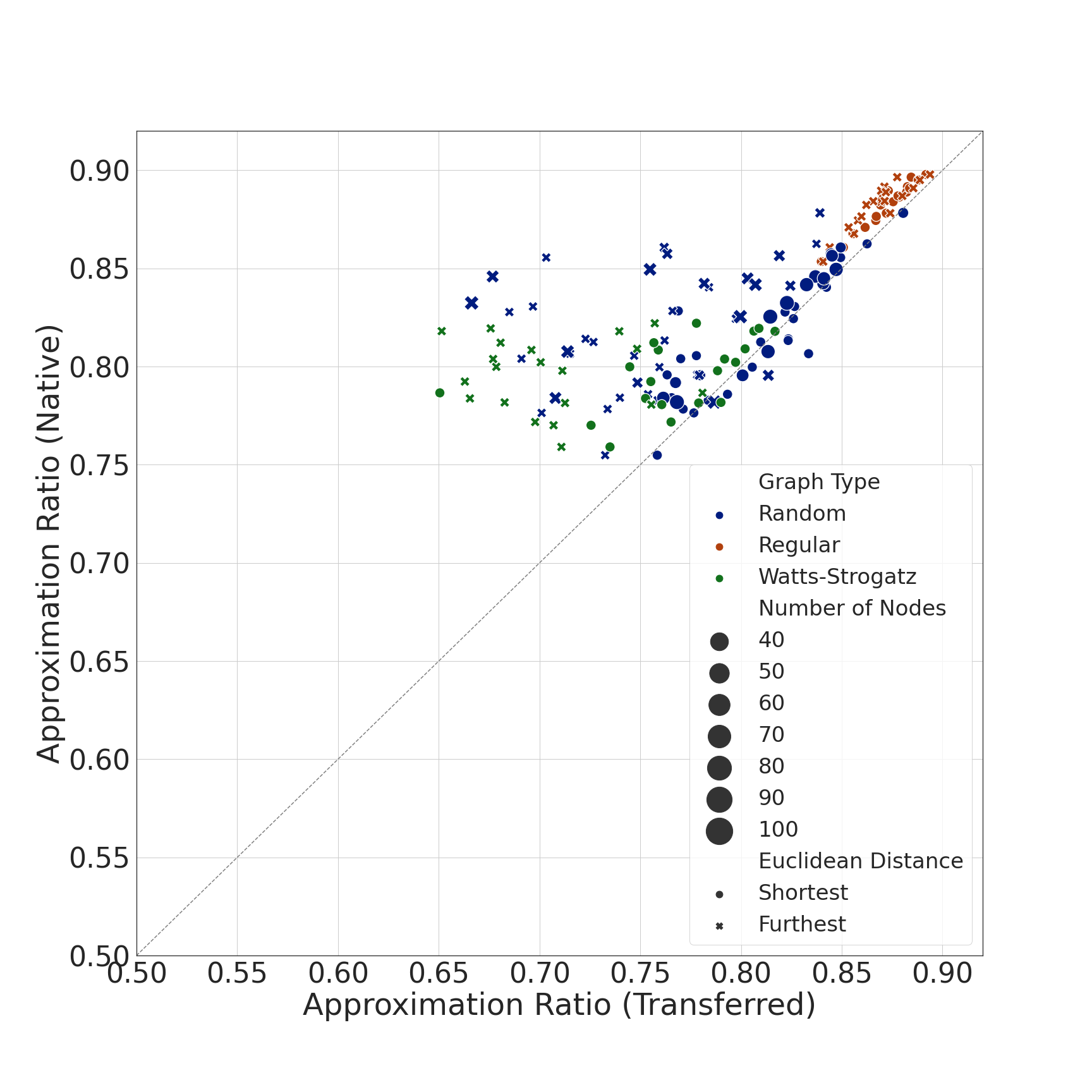}
    \caption{Transferability of optimal QAOA parameters with donor graphs chosen via Graph2Vec embedding. The data in circles show parameter transferability from donor graphs with the shortest Euclidean distance in embedded space to the acceptor graph. The data in crosses show parameter transferability from donor graphs with the furthest Euclidean distance in embedded space to the acceptor graph. Parameter transferability is also performed for random acceptor graphs of larger sizes (50-100 nodes) and for different classes of graphs (regular graphs and Watts-Strogatz graphs), shown in different colors.}
    \label{fig:Graph2Vec}
\end{figure}

Figure \ref{fig:Graph2Vec} shows parameter transferability for a set of acceptor graphs with donor graphs chosen via the Graph2Vec embedding technique. The transferred approximation ratio (Equation \ref{eq:avg_approx_ratio}) is compared against the native approximation ratio, computed as:
\begin{equation}
    \label{eq:approx_ratio}
    r^{*} = \frac{A(\Vec{\gamma}_{A^{*}},\Vec{\beta}_{A^{*}})}{A^{*}},
\end{equation}
where $A(\Vec{\gamma}_{A^{*}},\Vec{\beta}_{A^{*}})$ is the optimal QAOA energy for graph $A$. For comparison, parameter transferability is also performed for ``worst case scenarios'' by choosing donor graphs that have the largest (furthest) Euclidean distance from the test graphs (orange in Figure \ref{fig:Graph2Vec}). In general, the model is able to predict good donor candidates, particularly for similar MaxCut instances (same types of graphs), independent of acceptor graph size. To a lesser extent the model is able to predict good donor candidates for different types of MaxCut instances (transferability from random graphs to Watts-Strogatz graphs). In particular, for regular graphs\footnote{In general, the approximation ratios for regular graphs is greater than that of random and Watts-Strogatz graphs due to the fact that for regular graphs, a better approximate optimal solution is achieved for the same QAOA depth.}, the difference between ``best'' and ``worst'' donor candidates is not clearly demarcated.

\subsubsection{GL2Vec}

\begin{figure}[h!]
    \centering
    \includegraphics[scale=0.25]{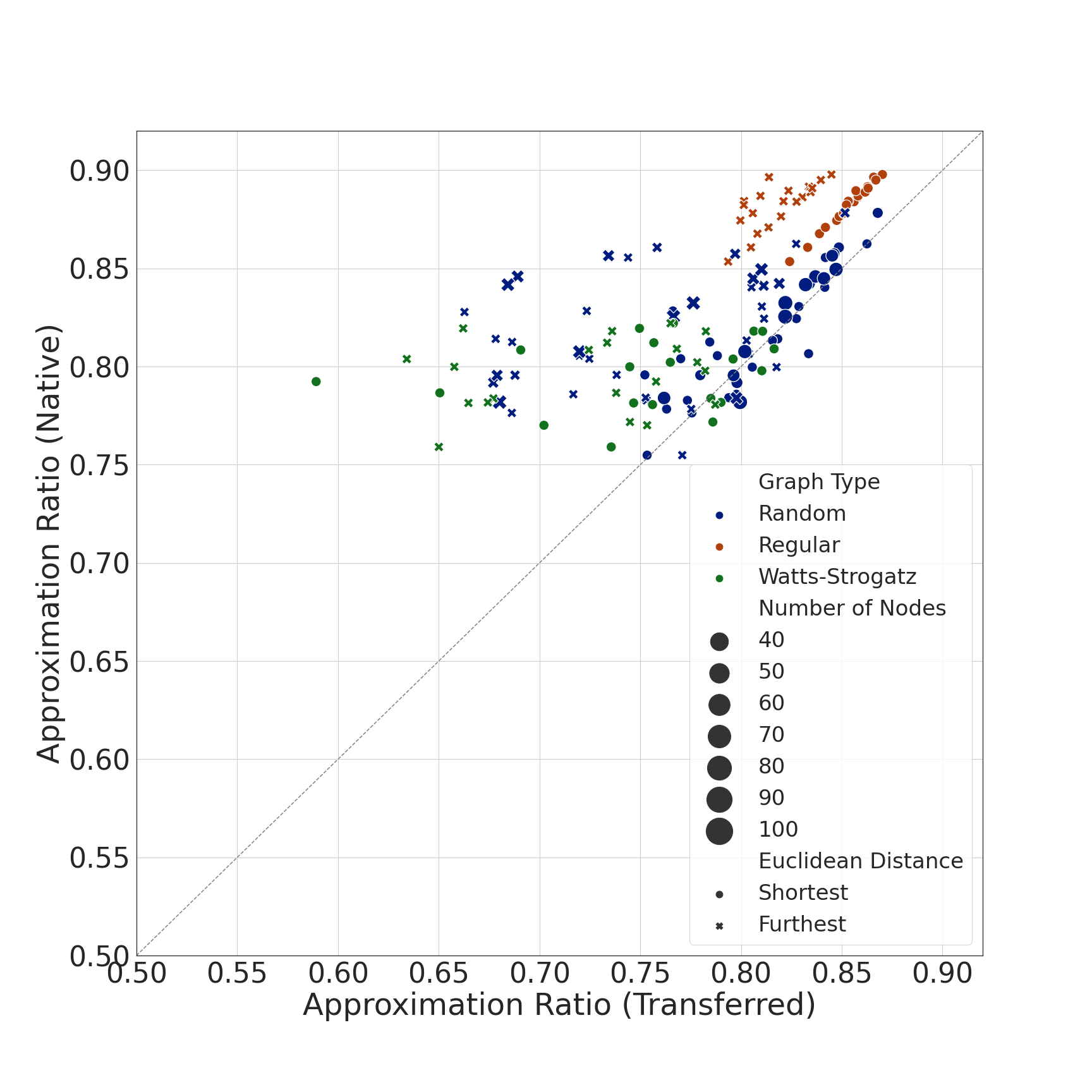}
    \caption{Transferability of optimal QAOA parameters with donor graphs chosen via GL2Vec embedding. Data in circles and crosses show shortest and furthest Euclidean distance in the embedded space, respectively. Different types of graphs are differentiated by color.}
    \label{fig:gl2vec}
\end{figure}

Figure \ref{fig:gl2vec} shows the approximation ratios for donor graphs chosen for target acceptor instances using the GL2Vec embedding model. The results for GL2Vec are fairly similar to those obtained with Graph2Vec embedding. Since the nodes and edges of our graphs do not contain any features, in the context of this work, The key difference between Graph2Vec and GL2Vec is that the latter 
overcomes some of the limitations by exploiting the dual graphs of given graphs. The GL2Vec performs slightly worse for regular graphs, yet is still able to differentiate between good donor candidates and poor donor candidates.

\subsubsection{Wavelet Characteristic}

\begin{figure}[h!]
    \centering
    \includegraphics[scale=0.25]{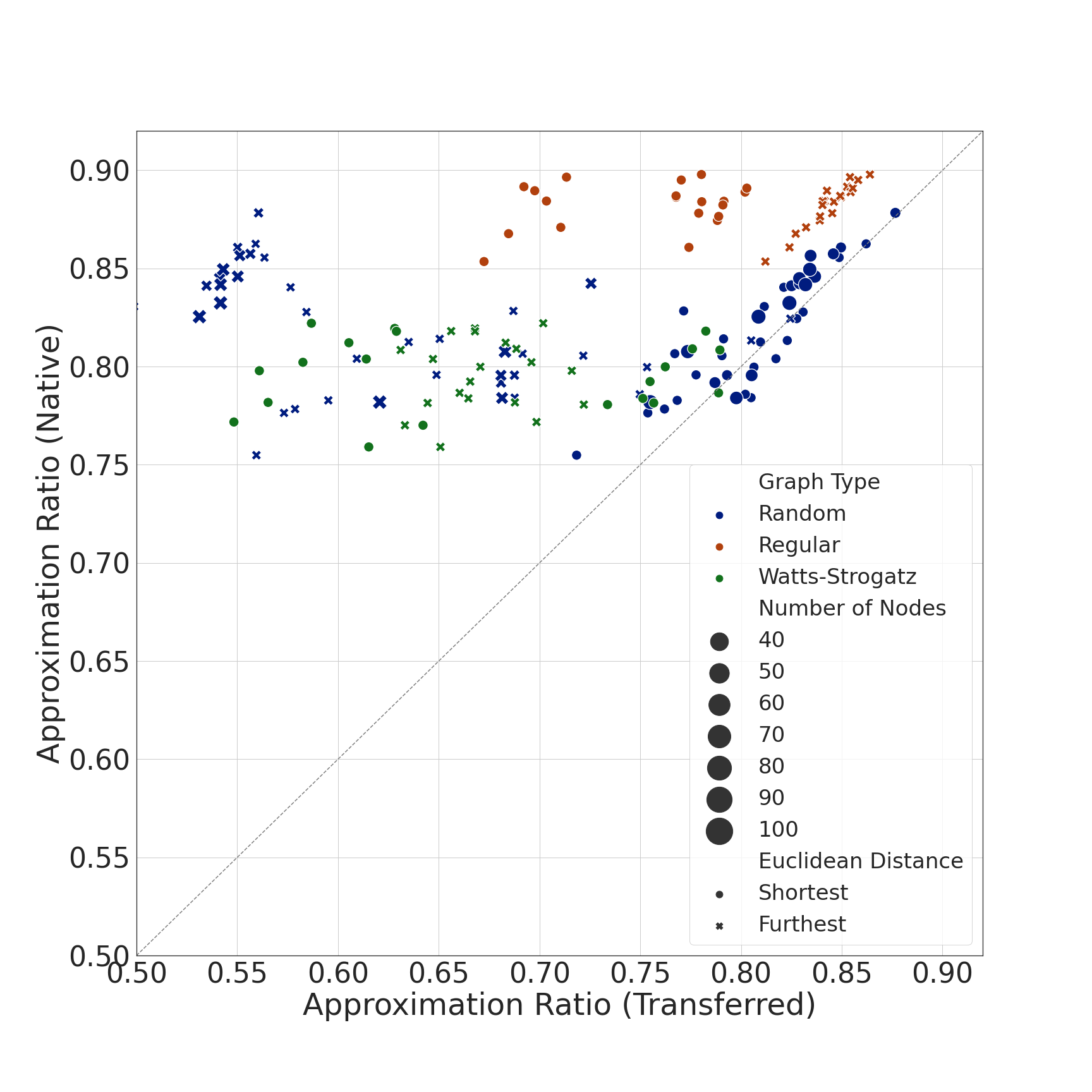}
    \caption{Transferability of optimal QAOA parameters with donor graphs chosen via Wavelet Characteristic embedding. Data in circles and crosses show shortest and furthest Euclidean distance in the embedded space, respectively. Different types of graphs are differentiated by color.}
    \label{fig:wavelet}
\end{figure}

The diffusion wavelets based graph embedding is the first of the embeddings we study that uses the Laplacian matrix of the graphs to perform graph level embedding. Figure \ref{fig:wavelet} shows the performance of the wavelet characteristic method for parameter transferability. In general, this method performs well for random acceptor graphs, including those with greater number of nodes. For regular graphs, this method performs poorly, as it predicts the opposite expected results: there is a higher transferred approximation ratio for those graphs that are further away in the embedded space. As for the Watts-Strogatz graphs, there seems to be no clear predictions or trends.

\subsubsection{Spectral Feature (SF)}

The SF embedding algorithm, like the wavelet characteristic embedding, uses the Laplacian matrix of graphs to perform embeddings, where the $k$ lowest eigenvalues of the Laplacian matrix are used as input for a classifier (in this case, a random forest classifier), that classifies graphs into types.

\begin{figure}[h!]
    \centering
    \includegraphics[scale=0.25]{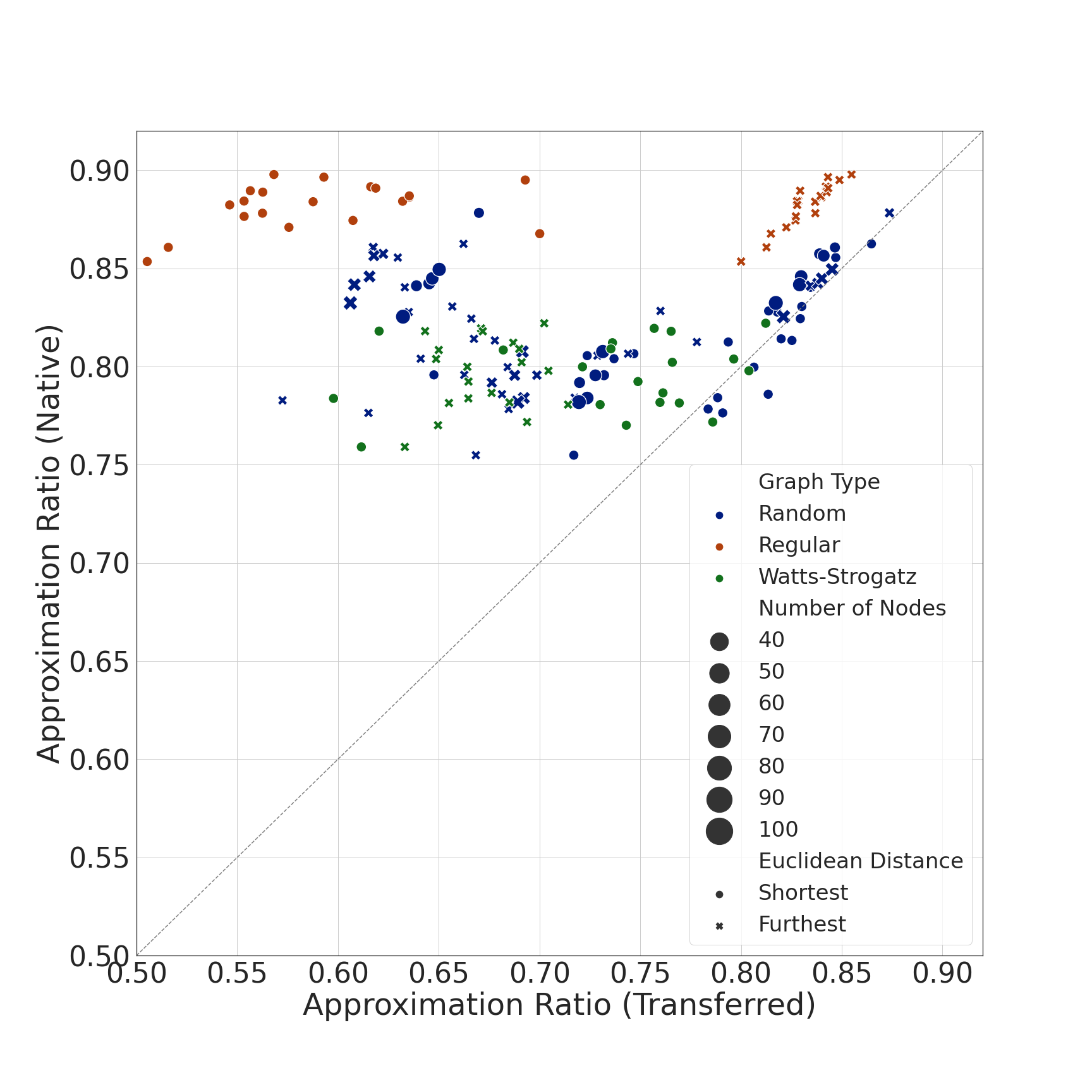}
    \caption{Transferability of optimal QAOA parameters with donor graphs chosen via SF embedding. Data in circles and crosses show shortest and furthest Euclidean distance in the embedded space, respectively. Different types of graphs are differentiated by color.}
    \label{fig:sf}
\end{figure}

Figure \ref{fig:sf} show the results for parameter transferability using the SF algorithm to predict good donor candidates. This embedding model does not predict good donor candidates, even for graphs that are the same type and have the same number of nodes as the graphs in the training set. In particular, SF performs poor predictions for regular graphs.

\subsubsection{FEATHER}

The FEATHER algorithm is based on characteristic functions defined on graph vertices to describe the distribution of vertex features at multiple scales. Specifically, the FEATHER algorithm calculates a specific variant of this characteristic functions (the r-scale random walk weighted characteristic function) where the probability weights of the characteristic function are defined as the transition probabilities of random walks. 

\begin{figure}[h!]
    \centering
    \includegraphics[scale=0.25]{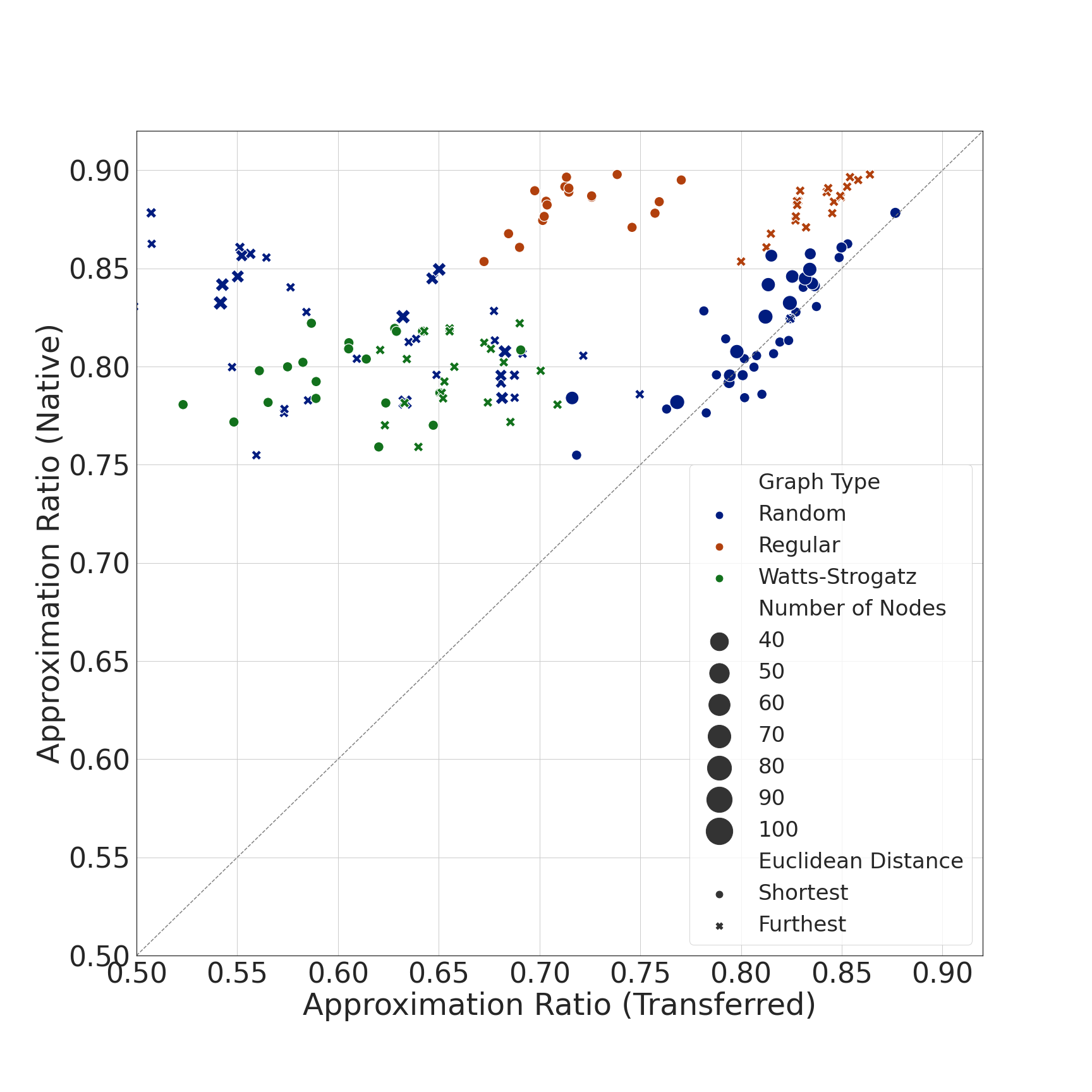}
    \caption{Transferability of optimal QAOA parameters with donor graphs chosen via FEATHER embedding. Data in circles and crosses show shortest and furthest Euclidean distance in the embedded space, respectively. Different types of graphs are differentiated by color.}
    \label{fig:feather}
\end{figure}

Figure \ref{fig:feather} shows the transferability results using the FEATHER embedding technique for donor graph prediction. In general, this method works well for random graphs; particularly those that are the same number of nodes as the graphs in the training set. As was the case for the wavelet characteristic and SF embeddings, this method does not provide good predictions for regular graphs. Furthermore, it predicts poorly for Watts-Strogatz acceptor graphs, as seen by the general trend of donor graphs closer in Euclidean distance giving lower transferred approximation ratios.

\subsubsection{ELRUNA}

Finally, for comparison purposes, we use the non-embedding, network alignment method ELRUNA~\citep{elruna} as a tool to predict good donor graph candidates for target acceptor graphs. With a network alignment technique, one can infer the similarities between cross-network vertices and discover potential node-level correspondence. ELRUNA relies exclusively on the underlying graph structure, computing the similarity between a pair of cross-network vertices iteratively by accumulating the similarities between their selected neighbors.

\begin{figure}[ht]
    \centering
    \includegraphics[scale=0.25]{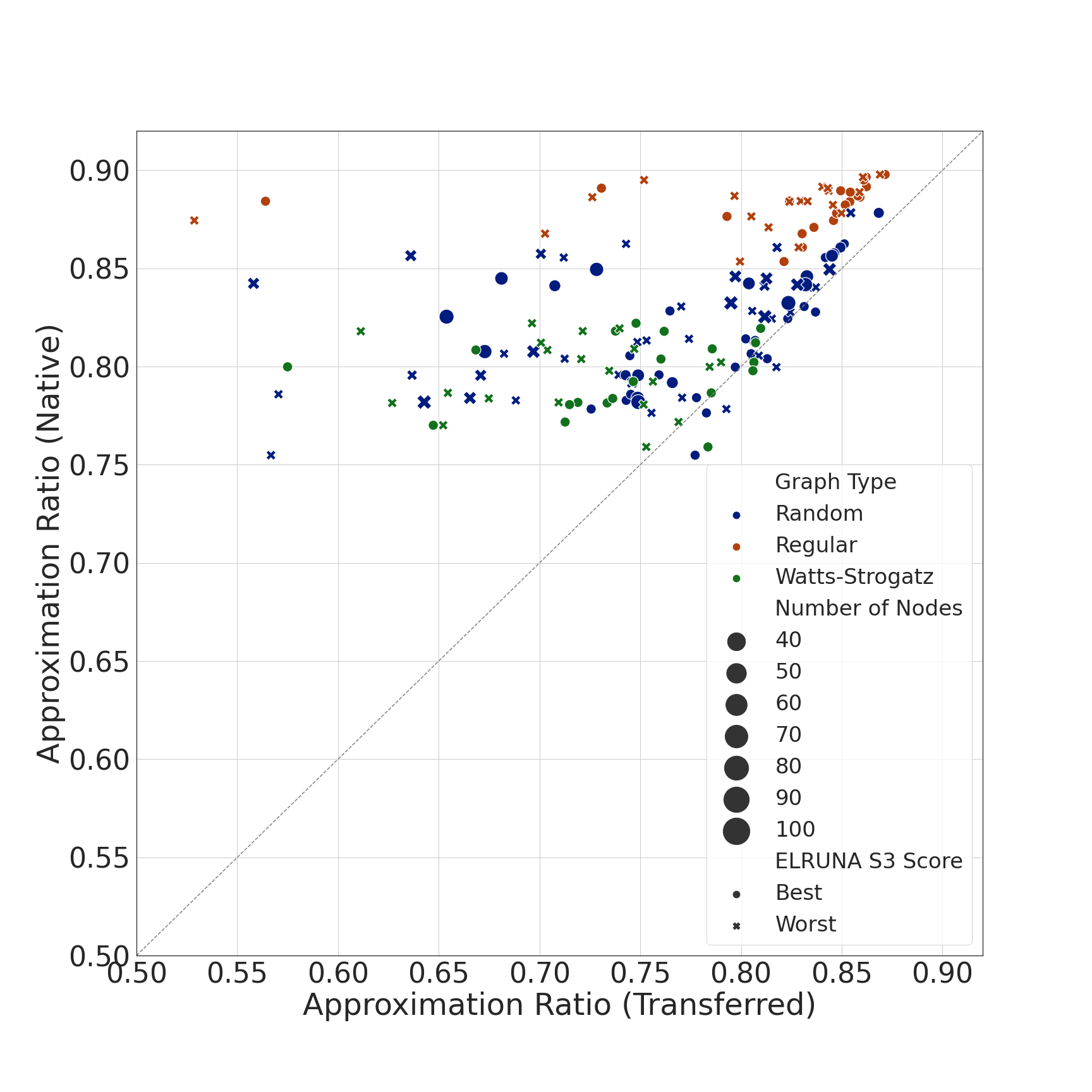}
    \caption{Transferability of optimal QAOA parameters with donor graphs chosen via ELRUNA. Data in circles and crosses show best and worst ELRUNA $S^{3}$ scores, respectively. Different types of graphs are differentiated by color.}
    \label{fig:elruna}
\end{figure}

Figure \ref{fig:elruna} shows the results for parameter transferability using ELRUNA's symmetric substructure score ($S^{3}$) as the predictor of good transferability. For isomorphic graphs, the score for $S^{3} = 1$; the higher the score, the more similar the graphs. As seen in the figure, ELRUNA does not predict good donor graph candidates reliably for any type of graph.

\subsection{Parameter Transferability for Larger QAOA Depth}

As previously mentioned, one of the features of a graph that is used to predict good optimal parameter transferability between two instances is the subgraph composition (light cones) between instances. As the depth of the circuit is increased, so does the size of the light cones. For large graphs, this does not pose an issue, as increasing the QAOA depth for large graphs becomes computationally intractable. Nevertheless, we study the effect of performing optimal parameter transferability for higher QAOA depth on smaller graph instances to determine whether parameter transferability is still effective. We begin by optimizing a training set of 4,500 20-node random graphs for a QAOA depth of $p = 10$\footnote{We decrease the graph sizes for the training set, as optimizing 40-node random graphs for a depth of $p = 10$ is not possible using a tensor network approach} and subsequently using Graph2Vec to generate the embedding for these graphs. The training set graphs were optimized using Cirq's state vector simulator for 10 seeds each. This model was tested with 22 20-node random graphs, 10 20-node 3-regular graphs, 10 20-node 4-regular graphs, and 20 Watts-Strogatz graphs. As before, optimal parameter transferability was performed for both best case (shortest Euclidean distance in embedded space) and worst case (furthest Euclidean distance in embedded space) scenarios.

\begin{figure}[ht]
    \centering
    \includegraphics[scale=0.25]{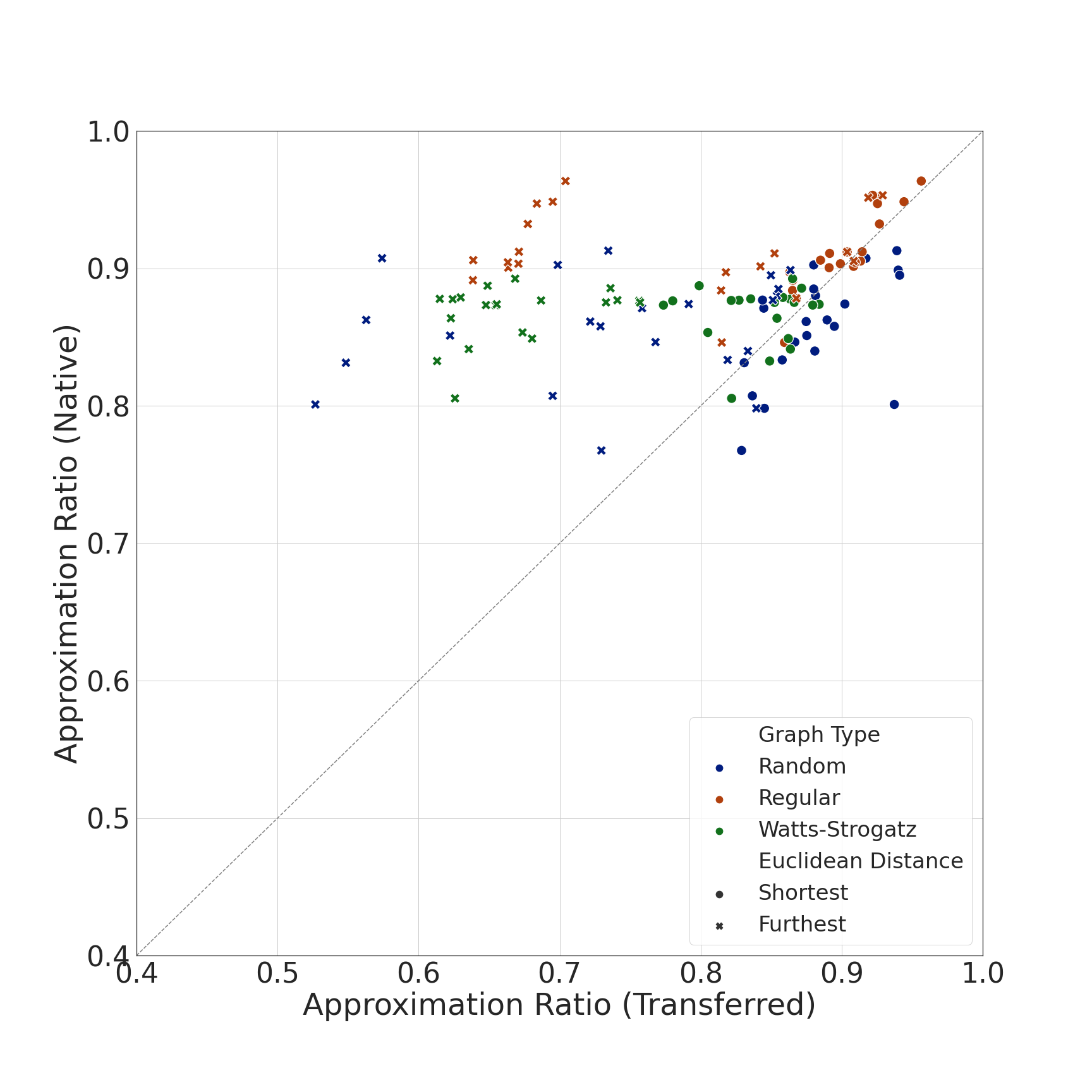}
    \caption{Optimal parameter transferability for a QAOA depth of $p = 10$ with donor graphs chosen via Graph2Vec embedding. All graphs in the test set are 20-node and different graph types are represented by different colors. Data in circles and crosses represent shortest and furthest Euclidean distance in embedded space, respectively.}
    \label{fig:state_vector_results}
\end{figure}

Figure \ref{fig:state_vector_results} shows the results obtained for parameter transferability using Graph2Vec for a QAOA depth of $p = 10$. We see in this figure that optimal parameter transferability still performs very  well for a higher QAOA depth. In most of the instances, the transferred approximation ratios from graphs that are closest in Euclidean distance are very close to the native approximation ratios, as opposed to the graphs that are furthest in Euclidean distance. As before, we see that there is a difference in performance for different types of graphs, where Graph2Vec seems to perform better for random and regular graphs than for Watts-Strogatz graphs. Notably, we see that we are able to perform optimal graph transferability effectively even though the Graph2Vec model is training on 4,500 graphs, as opposed to the previous (tensor network simulation) results, where the training state consisted of 30,000 graphs.

\subsection{Computational Time}

We test the computational performance of the Graph2Vec model for parameter transferability by running state vector simulations and comparing the MaxCut solution. These simulations were performed using Qiskit's state vector simulator on three 20-node graph instances, using a COBYLA optimizer for five different cases:

\begin{itemize}
\item QAOA optimization for 1,000 iterations starting with randomly initialized parameters\footnote{In all three instances, the optimizer reached convergence before 1,000 optimization steps}.
\item QAOA optimization for 100 iterations starting with randomly initialized parameters.
\item QAOA optimization for 10 iterations starting with randomly initialized parameters.
\item QAOA optimization for 10 iterations starting with transferred parameters obtained through the Graph2Vec model.
\item QAOA evaluation with transferred parameters obtained through the Graph2Vec model. In this case, no further optimization was applied.
\end{itemize}

\begin{figure}[ht]
    \centering
    \includegraphics[scale=0.6]{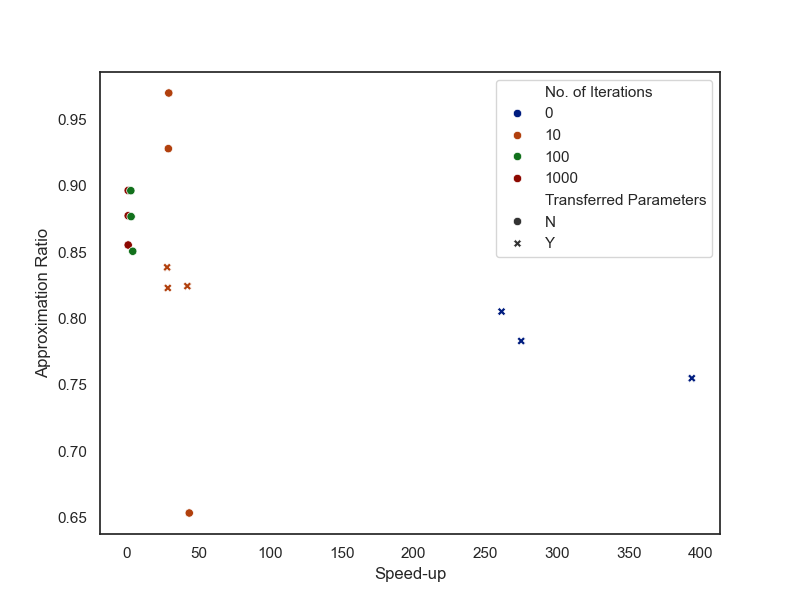}
    \caption{Computational speed up using Graph2Vec for parameter transferability in state vector simulations. The vertical axis shows the approximation ratio between the optimal solution and the solution obtained from state vector simulations. Circles represent no parameter transferability and the coloring represents how many optimization iterations were performed. Crosses represent parameter transferability using Graph2Vec and the coloring represents optimization steps.}
    \label{fig:sv_results}
\end{figure}

Figure \ref{fig:sv_results} shows the computational speed up afforded by parameter transferability using Graph2Vec. As can be seen in the figure, there is an interplay between parameter transferability and the approximation ratio. In the best case scenarios, one can expect to obtain a speed up of 200x-400x by transferring the parameters and evaluating the QAOA circuit (blue crosses in the figure), while still obtaining a similar approximation ratio. Optimizing a 20-node graph instance for 1,000 optimization steps takes 3-4 hours (red circles in the figure), while evaluation of a QAOA instance (with transferred parameters) takes $\sim$40 seconds. If, in addition to transferring optimal parameters, a few optimization steps are performed, it is possible to obtain better approximation ratios while still ensuring a $\sim$50x speedup.

\begin{table}
\begin{center}
\begin{tabular}{|c|c|c|c|c|c|}
\hline
    Graph Index & No. of Iterations & Transferred Parameters & Time (s) & Speed-Up & Approximation Ratio  \\
    \hline
    \multirow{5}{4em}{0} & 1000 & No & 16835.68 & 1.00 & 0.85 \\
    & 100 & No & 4036.62 & 4.17 & 0.85 \\
    & 10 & No & 386.35 & 43.58 & 0.65 \\
    & 10 & Yes & 398.56 & 42.24 & 0.83 \\
    & 0 & Yes & 42.75 & 393.85 &  0.75\\
    \hline
    \multirow{5}{4em}{1} & 1000 & No & 10921.13 & 1.00 & 0.89 \\
    & 100 & No & 3837.47 & 2.85 & 0.89 \\
    & 10 & No & 376.49 & 29.01 & 0.92 \\
    & 10 & Yes & 388.43 & 28.12 & 0.83 \\
    & 0 & Yes & 41.80 & 261.24 & 0.80 \\
    \hline
    \multirow{5}{4em}{2} & 1000 & No & 11534.55 & 1.00 & 0.87 \\
    & 100 & No & 3809.68 & 3.03 & 0.87 \\
    & 10 & No & 393.90 & 29.28 & 0.96 \\
    & 10 & Yes & 403.12 & 28.61 & 0.82 \\
    & 0 & Yes & 41.95 & 274.93 & 0.78 \\
    \hline
\end{tabular}
\end{center}
\caption{Summary of state vector simulation results to determine speed up from parameter transferability in 3 20-node graph instances.}
\label{tab:sv_results}
\end{table}

Table \ref{tab:sv_results} summarizes the state vector simulation results seen in Figure \ref{fig:sv_results}. We can observe that it is unnecessary to perform QAOA optimization for more than 100 iterations steps with randomly initialized parameters, as the optimal solution is obtained within this number of iterations, given the small graph instance. However, the optimal solution is not known in advance and the optimizer may require many more iterations for the convergence condition satisfaction. On the other hand, performing 10 optimization iterations on randomly initialized parameters is not enough to ensure a good solution to the problem (though some of the instances afford a better approximation ratio than those with optimized parameters, this is not always the case). Furthermore, if comparing the optimal cut (not shown in table) for all three cases, the optimal solution is obtained with optimal parameter transferability without the need for further optimization, so that, as a lower limit, at least a 100x speed up is guaranteed: comparing instances where 100 optimization steps are performed on randomly initialized parameters with those where a single evaluation of the QAOA circuit is performed with transferred optimal parameters.

\subsection{Results with Noise}

We simulated the performance of acceptor graphs on two IBM mock backends—the 14-qubit Guadalupe and the 27-qubit Auckland processors—using optimized QAOA parameters from donor graphs, which the graph embedding algorithm from Graph2Vec identified as having optimal transferability. The acceptor graphs are randomly generated 14-node graphs, each with a maximum degree of four per node. These simulations aim to assess the efficacy of the transferred parameters in scenarios reflective of real-world quantum hardware noise.

For both the Guadalupe and Auckland noise models, we simulated 1000 distinct 14-node graphs to compare their ideal, $E_{\text{ideal}}$, and noisy, $E_{\text{noisy}}$, expected energy values. We define the absolute error as $\Delta E = | E_{\text{noisy}} | - | E_{\text{ideal}} |$ and the relative error as $\Delta E/|E_{\text{ideal}}|$. Additionally, to reflect the continuous and dynamic nature of real processors, we repeated the experiments for the Gudalupe and Auckland processors using scale factors of $0.5$ and $2.0$ applied to the error rates in the corresponding fake backend noise models. This results in a total of six experiments: Gudalupe with scale factors of $0.5$, $1$, and $2.0$, and the same scale factors for Auckland.

\begin{figure}[h!]
    \centering
    \begin{subfigure}[b]{0.45\textwidth}
        \includegraphics[width=\textwidth]{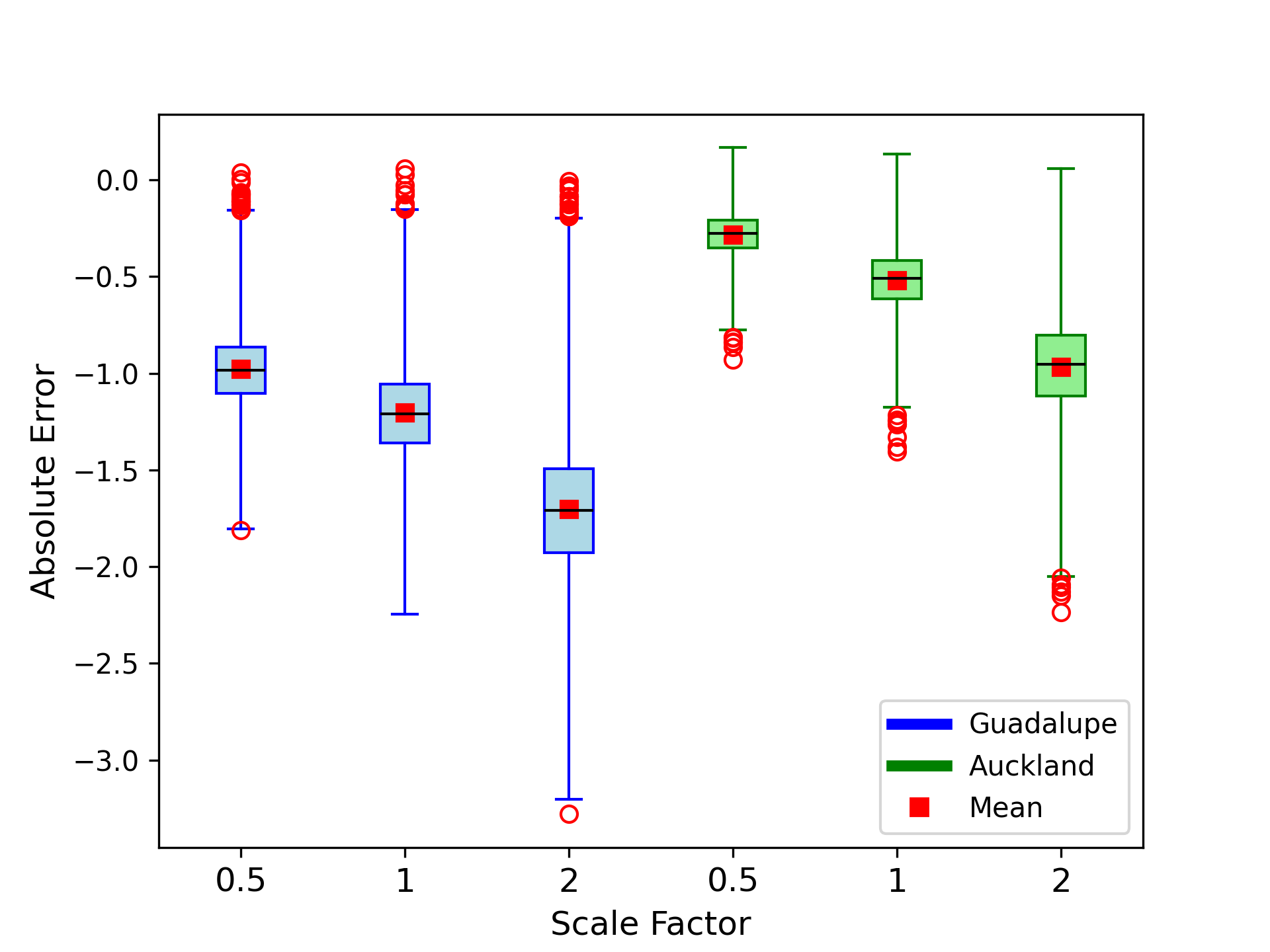}
        \caption{Absolute Error in Energy}
        \label{fig:absolute_error}
    \end{subfigure}
    \hfill
    \begin{subfigure}[b]{0.45\textwidth}
        \includegraphics[width=\textwidth]{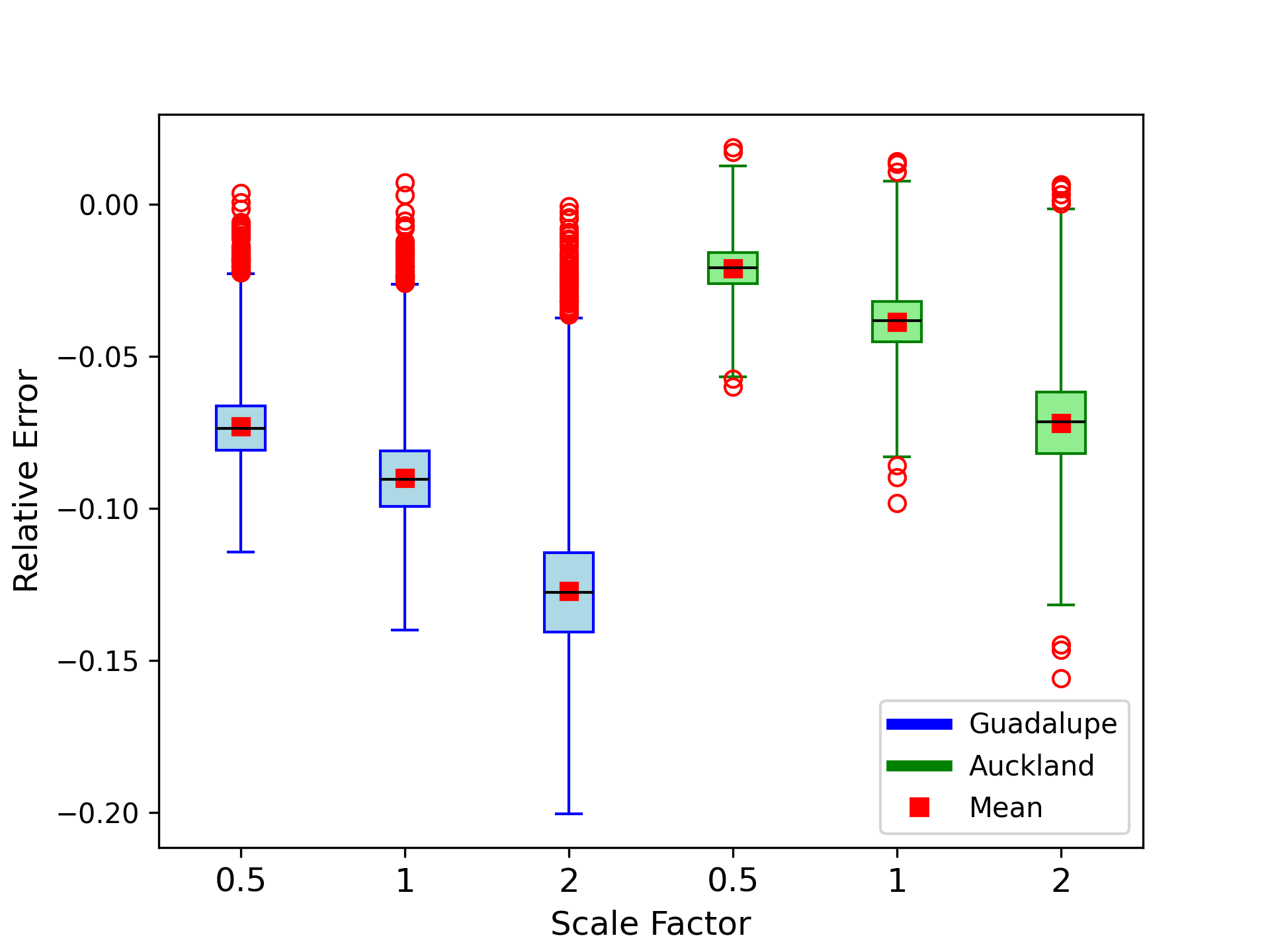}
        \caption{Relative Error in Energy}
        \label{fig:relative_error}
    \end{subfigure}
    \caption{Box plots depict (a) absolute and (b) relative errors in expected energy from QAOA simulations on the IBM mock backends: Guadalupe (blue) and Auckland (green). The horizontal axis represents the constant scaling factor we apply to the fake backend's error rates, with a scale factor of $1$ being the original error rates. The boxes capture the middle 50$\%$ of the data. Within each box, the solid line signifies the median, while the red square marker indicates the mean error. Data points outside the whiskers are outliers, shown as individual red circles.}
    \label{fig:boxplots}
\end{figure}

Figure \ref{fig:boxplots} illustrates the distribution of absolute and relative errors in comparison to the noiseless values. The median and mean errors suggest a modest deviation from the ideal for each scenario. There is also a clear trend as we vary the scale factor for both the Gudalupe and Auckland processors: larger error rates correspond to increased variability and error, which aligns with expectations. The Gudalupe processor appears to perform worse than the Auckland processor across all scale factors, yet the mean absolute (relative) error does not exceed $-2$ ($-0.15$). Additionally, the inter-quartile range, representing the middle $50\%$ of the data, shows that the ideal energy generally exceeds the noisy energy. This trend suggests a typical loss equivalent to one cut for the QAOA max cut problem on a 14-node graph. Overall, the relatively small discrepancy between noisy and ideal conditions suggests that the parameters transferred from donor graphs hold up effectively when applied to acceptor graphs in realistic quantum hardware environments.

\section{Discussion}\label{sec:discussion}

The results for parameter transferability, as measured by the transferred approximation ratio, show the power of using graph embedding techniques to determine good donor graph candidates for target acceptor graph instances. Of the five whole graph embedding models, and the one network alignment algorithm, it is clear that certain models are better donor graph predictors, in the context of optimal QAOA parameter transferability from random graphs in the MaxCut problem.

The first two graph embedding methods, Graph2Vec and GL2Vec, rely heavily on graph structural features for the learning procedure, with the difference that GL2Vec includes edge features as part of the learning procedure. Specifically, both of these learning methods create a \textit{corpus} of rooted subgraphs to train a model based on minimizing the log-likelihood probability of finding a particular subgraph in the context of other subgraphs. As mentioned in Section \ref{sec:graph_learning}, one of two graph features that predicts good transferability between to MaxCut instances is subgraph composition. Therefore, when a target acceptor graph is projected into the embedded space, its embedded vector will depend heavily on its subgraph composition. That is why, in general, we see good donor graph predictions for parameter transferability when using these types of embedding models. In particular, Graph2Vec predicts good donor candidates for most test graphs, including test graphs that are a different type to those in the training set. Furthermore, since we are applying QAOA to the unweighted MaxCut problem, there are no edge features, like edge weights, to take into consideration. For this reason, we see that there is no upside to using the GL2Vec model over the Graph2Vec model. This is reflected in the slightly worse performance of GL2Vec at determining good donor graph candidates for transferability.

As we move from whole graph embedding procedures to learning models that rely on the spectral features of graphs, namely, the SF and Wavelet Characteristic models, we see that these models do not offer an advantage to either Graph2Vec or GL2Vec for our particular use. Most pointedly, these two spectral features models perform poorly for regular graphs. This poor prediction for regular graphs can be attributed to the fact that the eigenvalues of the Laplacian matrix of regular graphs do not provide useful structural information, stemming from the degeneracy in values of the eigenvectors of the Laplacian. Interestingly, the SF algorithm is able to predict relatively well for Watts-Strogatz graphs. That is, for the entire test set of Watts-Strogatz graphs, the graphs that show a better approximation ratio had all vectors that were closer in Euclidean space to the training set's graph vectors, and those with worse approximation ratios had vectors that were further away in Euclidean space.

Even though embedding techniques based on structural graph features are not well suited for regular graphs, they can be used to embed weighted graphs, as a graph's weight information is encoded in its Laplacian. Therefore, spectral embedding techniques can be used to perform transferability of optimal parameters in weighted instances of graphs. This can be applied to a multi-level QAOA approach, greatly improving the time performance of weighted MaxCut QAOA optimization \citep{bach_mlqaoa_2024}.

The FEATHER graph embedding model, which uses the r-scale random walk weighted characteristic function to compute transition probabilities of random walks, again, does not offer an advantage to the structural graph features models when looking at predictions beyond random graphs. For random graphs, we see that this model predicts donor graph candidates fairly well, but it fails at predicting good donor candidates for regular graphs and Watts-Strogatz graphs. As was the case for the SF and Wavalet Characteristic algorithms, the FEATHER algorithm predicts better donor graph candidates that are \textit{further} away in the embedded space for regular acceptor graphs.

Finally, for comparison purposes, we show the predictions of donor graphs for target acceptors using the network alignment method ELRUNA. As seen from the results, this method does not offer an advantage to determining good donor graph candidates, even for the cases where both donor and acceptor graphs are of the same type (random) and have the same number of nodes.

Based on the results across all five whole graph embedding algorithms, the Graph2Vec algorithm proves to be the best overall algorithm at determining good donor graph candidates for target acceptor instances, including instances where the type of acceptor graph is different from the graph type in the training set of the model. Furthermore, we show that Graph2Vec can still be employed for optimal parameter transferability in graph instances that have been optimized for larger QAOA depths. For specific cases where the target acceptor graph is the same type as the graphs in the training set, an algorithm like FEATHER performs good predictions, although it is cannot be generalized to include different types of graphs.

Looking at the state vector simulations, the computational speed-up that our method affords points to the potential of using graph embedding techniques for parameter transferability, both as a warm-start for further optimization or for direct evaluation of the QAOA circuit with optimal transferred parameters.

Incorporating noise into our evaluation, Graph2Vec showed that parameters from donor graphs can still perform effectively on acceptor graphs, with only a minor deviation from ideal performance. This demonstrates the algorithm's potential for parameter transfer in quantum processors similar to IBM's Guadalupe and Auckland, affirming its applicability in the NISQ era.

\section{Conclusion}\label{sec:conclusions}

In this work, we compared five whole graph embedding algorithms and one network alignment algorithm for the task of finding a good donor graph candidate for an instance of optimal QAOA parameter transferability in the MaxCut problem. Among the set of five whole graph embedding algorithms, the Graph2Vec algorithm had the best overall performance, with the FEATHER algorithm also performing well for instances where the training set graphs and the target test graph are of the same type.

All of these methods could be potentially improved by including different types of graphs in the training of the models, a topic that can be addressed in future work. In particular, whole graph embedding model like SF can see an improvement with the addition of different types of graphs in the training set as it performs predictions based on a classifier.

Furthermore, the graph embedding techniques that rely on structural graph features, Graph2Vec and GL2Vec, can be improved by including additional graph features into the model. Ideally, these would include features that predict successful optimal parameter transferability between two QAOA instances for the MaxCut problem, much like subgraph composition and parity do. As of the writing of this article, there are no other well-studied graph features that predict good QAOA parameter transferability between two MaxCut instances. Our current research efforts are aimed at finding additional features that predict good transferability between two graph instances, and more broadly understanding rigorously why subgraph composition and parity of graphs are good predictors of parameter transferability.

Given the noise inherent in modern quantum processors, our results demonstrate that parameters obtained from graph embeddings retain their utility. The resilience of these parameters against noise points to their potential for improving QAOA during the NISQ era. Incorporating error mitigation techniques, such as Zero-Noise Extrapolation (ZNE), should further enhance this robustness \citep{cai2023quantum}. Consequently, our findings demonstrate the viability of parameter transferability as a practical approach for accelerating QAOA.

As a whole, employing a graph embedding technique, such as Graph2Vec, to select optimal QAOA parameters for transferability greatly reduces the computational cost of performing a native optimization procedure on a graph instance. On the one hand, this approach can overcome the issue of encountering barren plateaus during the local optimization process, especially for the cases where noise is present. On the other hand, optimal parameters can be transferred to larger graph instances, reducing the computational time required to natively solve the large instance, as the cost increases with the size of the instance. In short, using a graph embedding approach for parameter transferability can greatly improve the computational cost associated with QAOA parameter optimization. 

Particularly interesting future research questions is: can we combine the parameter transferability with such techniques the QAOA sparsification \citep{liu2022quantum}, problem symmetry learning \citep{tsvelikhovskiy2023representation,shaydulin2021classical} and using different than uniform distributions for initialization \citep{kulshrestha2022beinit} to accelerate and make more robust QAOA even more?

\section*{Data Availability}
The code used to produce the results of this article can be found in the GitHub repository: \url{https://github.com/joseluisfalla/QPTransfer}.

\section*{Acknowledgements}

This work was supported  in part with funding from the Defense Advanced Research Projects Agency (DARPA). The views, opinions and/or  findings expressed are those of the authors and should not be interpreted as representing the official views or policies  of the Department of Defense or the U.S. Government.

This research was supported in part through the use of DARWIN computing system: DARWIN - A Resource for Computational and Data-intensive Research at the University of Delaware and in the Delaware Region, which is supported by NSF Grant \#1919839.

This research was partially supported by NSF, under award number 2122793.

This material is based upon work supported by the U.S. Department of Energy, Office of Science, National Quantum Information Science Research Centers. Y.A. acknowledges 
support from the U.S. Department of Energy, Ofﬁce of Science, under contract DE-AC02-06CH11357 at Argonne National Laboratory.

\newpage
\bibliographystyle{unsrtnat}
\bibliography{references}

\end{document}